\documentclass[aps,preprint, showpacs]{revtex4}
\input{psfig.sty}

\usepackage{amsmath,bm}
\newcommand{\beq}{\begin{eqnarray}}
\newcommand{\eeq}{\end{eqnarray}}

\newcommand{\s}{\bm{\sigma}}

\newcommand{\J}{\bm{J}}

\newcommand{\K}{\mathbf{K}}

\newcommand{\p}{\bm{p}}

\newcommand{\hp}{\bm{\widehat{\p}}}

\newcommand{\x}{\bm{x}}

\newcommand{\0}{\bm{0}}

\newcommand{\bv}{\bm{\varphi}}

\newcommand{\hbv}{\bm{\widehat\varphi}}

\newcommand{\bl}{\bm{\lambda}}

\newcommand{\br}{\bm{\rho}}

\begin{document}

\title{$N$ and $\Delta$ Resonances in the Rigid Quark--Di-Quark Vibrator}

\author{M.\ Kirchbach}

\affiliation{Instituto de F\'{\i}sica, \\
         Universidad Aut\'onoma de San Luis Potos\'{\i},\\
         Av. Manuel Nava 6, San Luis Potos\'{\i}, S.L.P. 78290, M\'exico}

\date{\today}

\begin{abstract}
A nearest-neighbor analysis of light flavor baryon  mass 
spectra reveals striking degeneracy patterns
-- narrow mass bands populated by series of 
parity twins of steadily increasing spins.
Each series terminates by an unpaired resonance 
of the highest spin in the group.
We trace back such degenerate series of resonances
(to be termed by us as ``mega states'') to internal baryon structure 
dominated by a quark--(rigid di-quark) configuration.
Nucleon and $\Delta $ spectra are, surprisingly enough,
exact replicas to each other in the sense that
each of them features three mega states of equal quantum numbers.
We fit positions and splittings of the observed mega states 
by an algebraic Hamiltonian which translates into
a potential that is a combination of
Coulomb-- and Morse-like  potentials 
and predict few more such degenerate series to be observed
by the TJNAF ``missing resonance'' program.
Finally, we explore consequences of the model
for some electrodynamic properties of the spectra.

\end{abstract}


\maketitle

\def\beq{\begin{eqnarray}}
\def\eeq{\end{eqnarray}}

\def\A{{\mathcal A}^\mu}
\def\W{{\mathcal W}_\mu}

\def\beq{\begin{eqnarray}}
\def\eeq{\end{eqnarray}}


\def\s{\mbox{\boldmath$\displaystyle\mathbf{\sigma}$}}
\def\S{\mbox{\boldmath$\displaystyle\mathbf{\Sigma}$}}
\def\J{\mbox{\boldmath$\displaystyle\mathbf{J}$}}
\def\K{\mbox{\boldmath$\displaystyle\mathbf{K}$}}
\def\P{\mbox{\boldmath$\displaystyle\mathbf{P}$}}
\def\p{\mbox{\boldmath$\displaystyle\mathbf{p}$}}
\def\hp{\mbox{\boldmath$\displaystyle\mathbf{\widehat{\p}}$}}
\def\x{\mbox{\boldmath$\displaystyle\mathbf{x}$}}
\def\0{\mbox{\boldmath$\displaystyle\mathbf{0}$}}
\def\bv{\mbox{\boldmath$\displaystyle\mathbf{\varphi}$}}
\def\hbv{\mbox{\boldmath$\displaystyle\mathbf{\widehat\varphi}$}}

\def\bl{\mbox{\boldmath$\displaystyle\mathbf{\lambda}$}}
\def\bl{\mbox{\boldmath$\displaystyle\mathbf{\lambda}$}}
\def\br{\mbox{\boldmath$\displaystyle\mathbf{\rho}$}}
\def\1{\openone}
\def\bfhh{\mbox{\boldmath$\displaystyle\mathbf{(1/2,0)\oplus(0,1/2)}\,\,$}}

\def\mn{\mbox{\boldmath$\displaystyle\mathbf{\nu}$}}
\def\amn{\mbox{\boldmath$\displaystyle\mathbf{\overline{\nu}}$}}

\def\mne{\mbox{\boldmath$\displaystyle\mathbf{\nu_e}$}}
\def\amne{\mbox{\boldmath$\displaystyle\mathbf{\overline{\nu}_e}$}}
\def\rlh{\mbox{\boldmath$\displaystyle\mathbf{\rightleftharpoons}$}}

\def\wm{\mbox{\boldmath$\displaystyle\mathbf{W^-}$}}
\def\hh{\mbox{\boldmath$\displaystyle\mathbf{(1/2,1/2)}$}}
\def\h00h{\mbox{\boldmath$\displaystyle\mathbf{(1/2,0)\oplus(0,1/2)}$}}
\def\znbb{\mbox{\boldmath$\displaystyle\mathbf{0\nu \beta\beta}$}}



\section{Order versus uniformity in light-quark baryon spectra.}
The importance of the spectrum of composite systems for unveiling 
dynamics and relevant degrees of freedom 
in the theories of the micro-world is hardly to be overestimated.
Suffices to recall that quantum mechanics was established only 
after the successful description of the experimentally observed 
degeneracy patterns of the Balmer series in the excitations of 
the hydrogen atom. 
Furthermore, in solid state physics, a spectrum with a continuous and 
gap-less branch of the low--lying excitations like that of the infinite
ferromagnet, has been decisive for unveiling the magnons as the relevant
degrees of freedom and for establishing the mechanism of the spontaneous
reduction of rotational symmetry. A further  high impact example
is a spectrum with a low lying gap, like the one of superconducting materials,
that hints onto spontaneous breaking of local symmetries and
signals massiveness of gauge fields.
It is not exaggerated to notice that spectra are Nature messengers 
about the dynamical properties of internal structure.

\vspace{0.15cm}
\noindent
{}For same reasons, search for degeneracy patterns of baryon 
excitations is interesting and instructive for uncovering the 
dynamical properties of the theory of strong interaction-- 
the Quantum Chromo- Dynamics.

\vspace{0.15cm}
\noindent
The structure of the nucleon spectrum is far from being settled
despite its long history.
The situation relates to the fact that the first facility
to measure $N$ and $\Delta $ levels, the Los Alamos Meson Physics Facility 
(LAMPF), ceased to encounter all the states that were possible as excitations 
of three quarks. Later on, 
the Thomas Jefferson National Accelerator Facility (TJNAF) was designed 
to search (among others) for those ``missing resonances''.
At present, data have been collected and are awaiting evaluation
\cite{Burkert}.  

\vspace{0.15cm}
\noindent
In the series of papers \cite{MK97-98} I performed a
nearest neighbor analysis of all available data 
~\cite{PART} on mass distribution of
nucleon resonances, and drew attention 
to the phenomenon of {\it state--density increase\/}  
in {\it three\/} particularly  {\it narrow\/} 
mass bands and its almost exact replica in the  
$\Delta $ spectrum (see Fig.~ 1). 

\vspace*{1truecm}
\begin{figure}[htb]
\vskip 5.0cm
\includegraphics{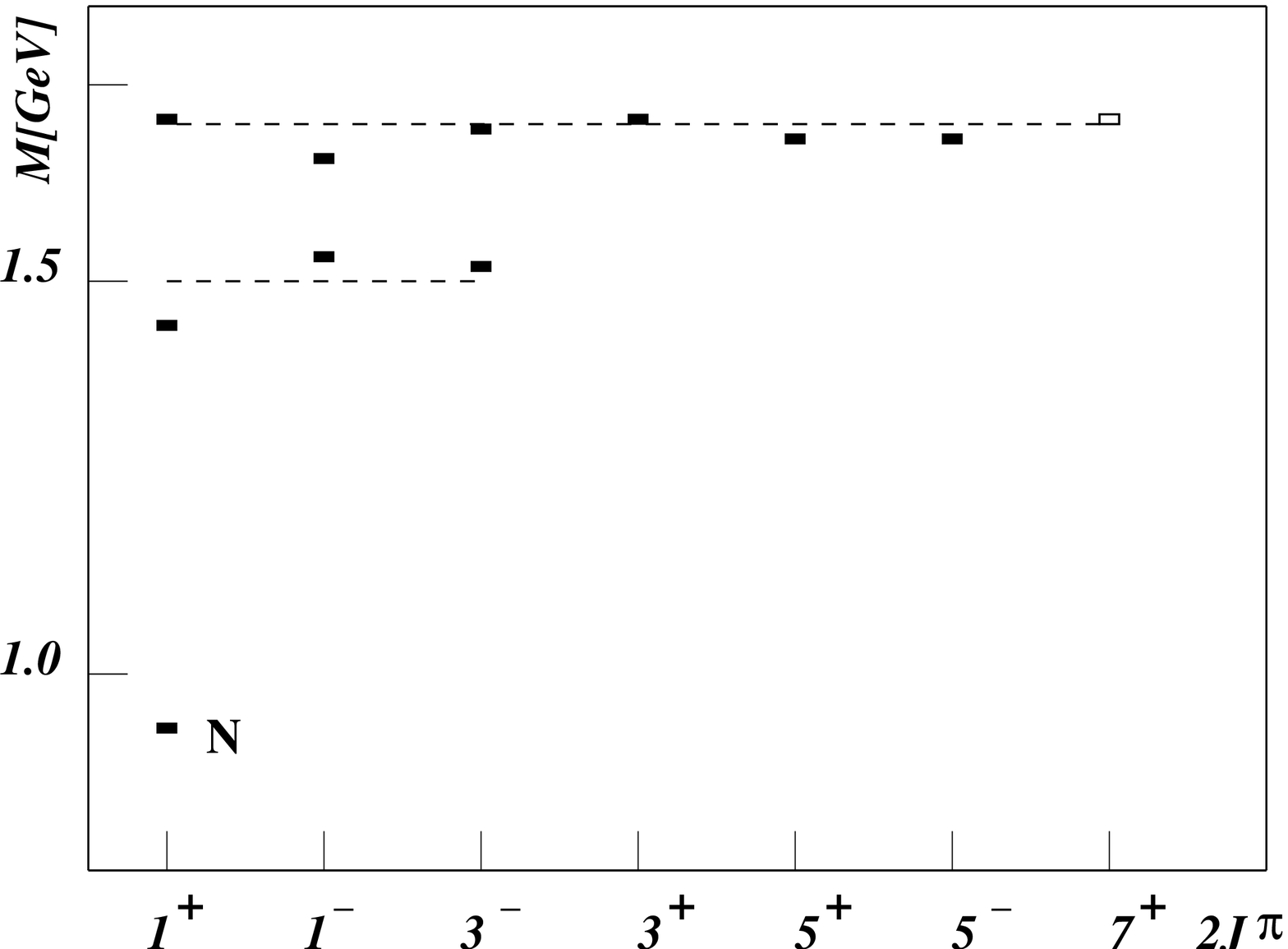}
\includegraphics{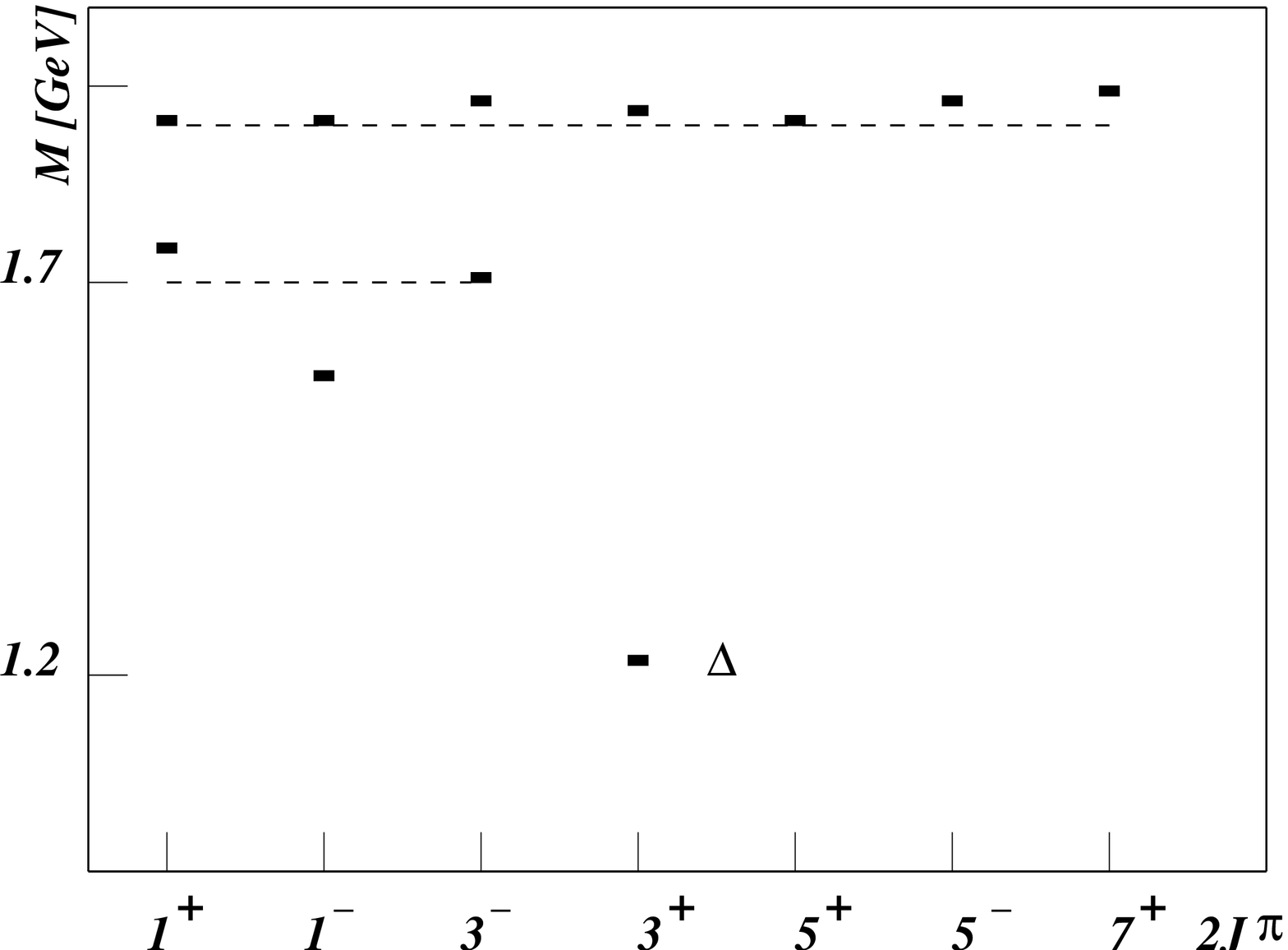}
\vspace{1.01cm}
{\small Fig.~1. 
Summary of the data on $N$ and $\Delta$ resonances with
masses below 2000 MeV (without the $\Delta (1600)$ state).
The filled bricks represent known resonances, while 
the sole empty brick in the $N$ spectrum
corresponds to the prediction of the $F_{17}(\sim 1700)$
resonance. See Ref.~\cite{MK_Evora} for the full spectra.
}
\end{figure}
\vspace{0.15cm}
\noindent
The first bunch of nucleon states consists of the parity 
twins ($P_{11}(1440)$--$S_{11}(1535)$), and
the edge $D_{13}(1520)$ resonance.
It is paralleled in the $\Delta $ spectrum by
($P_{31}(1750)$--$S_{31}(1620)$) and $D_{33}(1700)$, respectively.

\vspace{0.15cm}
\noindent
The second series of $\Delta$ resonances starts with the three parity twins 
($P_{31}(1910)$--$S_{31}(1900)$), ($P_{33}(1920)$--$D_{33}(1940)$),
and $(F_{35}(1905)$--$D_{35}(1930)$),
with spins ranging from ${1\over 2}^\pm$ to ${5\over 2}^\pm $,
and terminates with the edge spin-${7\over 2}^+$ resonance, 
$F_{37}(1950)$. In the nucleon spectrum this series is paralleled
by ($P_{11}(1710)$--$S_{11}(1650)$), ($P_{13}(1720)$--$D_{13}(1700)$),
and $(F_{15}(1680)$--$D_{15}(1675)$), while the
counterpart to $F_{37}$ is ``missing''.

\vspace{0.15cm}
\noindent
{}Finally, the third series consists of
five parity twins  with spins ranging
from ${1\over 2}^\pm$ to ${9\over 2}^\pm $
(see Refs.~\cite{PART}, \cite{MK_Evora}) 
and terminates by the edge  $H_{3, 11}(2420)$ resonance.
A comparison between the $N$ and $\Delta $ spectra shows that 
the third $N$ and $\Delta$ series 
are identical up to four more unoccupied resonances.
These are: (i) the nucleon  counterpart to 
four-star $\Delta$ resonances $H_{3, 11}(2420)$, and 
(ii) the $\Delta $ counterparts to the $N$ (one--and two star) 
resonances $P_{11}(2100)$, $P_{13}(1900)$, and $D_{13}(2080)$,
respectively. 
Above analysis suggests to {\it characterize\/} the series of 
(quasi)degenerate resonances by the {\it number\/} (here denoted by $K$), of the 
{\it parity twins\/} contained there. 
In the following we shall refer to such series as {\it $K$-mega states\/}.

\vspace{0.15cm}
\noindent
The $\Delta (1600)$ resonance in being
most probably an independent hybrid state, is the only state
that at present drops out of our systematics.
Compared to resonance classification schemes based upon 
$SU(6)_{sf}\times O(3)_L$, the light flavor baryon spectra clearly
reveal regularities rather than the expected uniformity.
 
\vspace{0.15cm}
\noindent
The existence of identical $N$ and $\Delta $ degeneracy patterns,
raises the question as to what extent are we here facing a new type of 
symmetry which was not anticipated by anyone of the market
models or theories.
The next Section devotes itself to answering this question.

\section{ Spectroscopy of mega states. }
 
\subsection{QCD motivated necessity for a quark--di-quark configuration.
A preliminary.}
To the extent QCD prescribes baryons to be constituted of three
quarks in a color singlet state, one can 
exploit for  the description of baryonic systems
algebraic models developed for the purposes of triatomic molecules,
a path pursued by  Refs.~\cite{U(7)}.
An interesting dynamical limit of the three quark system is
$U(7)\longrightarrow U(3)\times U(4)$, when two of the quarks 
act as an independent entity, a di-quark (Dq), 
while the third quark (q) remains spectator.
While the di-quark approximation~\cite{Torino} turned out to be rather 
convenient for describing various properties of the ground state 
baryons \cite{ReinA}, \cite{Kusaka},  applications to excitation
spectra are still lacking. It is one of our goals to partly fill this gap. 

\vspace{0.15cm}
\noindent
The usefulness, even necessity, for having a quark--di-quark 
configuration within the nucleon is independently supported by 
arguments related to spin in QCD. 
In particular, the meaning of spin in QCD was re-visited
in Refs.~\cite{Doug1}, and \cite{Doug2} in connection with the proton 
spin puzzle.

\vspace{0.15cm}
\noindent
As is well known, the spin degrees of freedom 
of the valence quarks alone are not
sufficient to explain the spin-${1\over 2}$ of the nucleon.
Rather, one needs to account for
the orbital angular momentum of the quarks 
(here denoted by $L_{QCD}$) and the angular momentum carried
by the gluons (so called field angular momentum, $G_{QCD}$):
\begin{eqnarray*}
{1\over 2} &=&{1\over 2}\Delta \Sigma  +L_{QCD}  +G_{QCD} \, \\
= \int d^3 x {\Big[} {1\over 2}\bar \psi \vec{\gamma}\gamma_5\psi
&+&\psi^\dagger (\vec{x}\times (-\vec{D}))\psi 
+\vec{x}\times (\vec{E}^a\times \vec{B}^a){\Big]}\, .
\end{eqnarray*}
In so doing one encounters the problem that neither $L_{QCD}$, nor $G_{QCD}$
satisfies the spin $su(2)$ algebra. If at least  
$\left(L_{QCD} +G_{QCD}\right)$ is to be given the interpretation
of angular momentum, then one has to require validity of the $su(2)$ 
commutator
\beq
{\Big[} 
\left( L_{QCD}^i +G^i_{QCD} \right),
\left( L_{QCD}^j +G^j_{QCD}\right) 
{\Big]}
=i
\epsilon^{ijk} \left( L_{QCD}^k +G^k_{QCD}\right)\, .
\label{su2_alg}
\eeq	
Equation (\ref{su2_alg}) restricts
$\vec{E}^{i;a}$ to a chromo-electric charge,
\beq
E^{i;a}={{gx' \,  ^i} \over {r'\, ^3}}T^a\, ,
\label{E-chromo}
\eeq
while $\vec{B}^{i;a} $ has to be a chromo-magnetic dipole according to, 
\beq
B^{ia}= ({ {3x^i \sum_l x^l m^l}\over r^5 } -
{m^i\over r^3})T^a\, ,
\label{B-chromo}
\eeq
where  $x'\, ^i=x^i-R^i$.
Above color fields correspond to the perturbative one-gluon 
approximation typical for a di-quark-quark structure. 
The di-quark and the quark are in
turn the sources of the color Coulomb--,
and the color magnetic--dipole fields.
In terms of color and flavor degrees of freedom,
the nucleon wave function does indeed have the required quark--di-quark
form 
\beq
\vert p_{\uparrow} \rangle =
{ \epsilon_{ijk} \over \sqrt{18} }\,
\lbrack u^+_{i \downarrow} d^+_{j\uparrow} -
u^+_{i\uparrow}d^+_{j\downarrow} \rbrack\,  u^+_{k\uparrow}\,\vert 0\rangle .
\label{q-Dq}
\eeq
A similar situation appears when looking up the covariant QCD solutions
in form of a membrane with the three open ends being associated with the 
valence quarks. When such a membrane stretches to a string (see Fig.~2), 
so that a linear action (so called gonihedric string) can be considered, 
one again encounters same $K$-mega state degeneracies in the excitations 
spectra of the baryons, this time as a part of an infinite $K$ tower.
The result was  reported by Savvidy in Ref.~\cite{Savvidy}.
Thus not only the covariant spin-description provides an independent argument
in favor of a dominant quark-di-quark configuration in the 
nucleon structure, but also search for covariant  
resonant QCD solutions leads once again to same picture. 
\begin{figure}[htbp]
\centerline{\psfig{figure=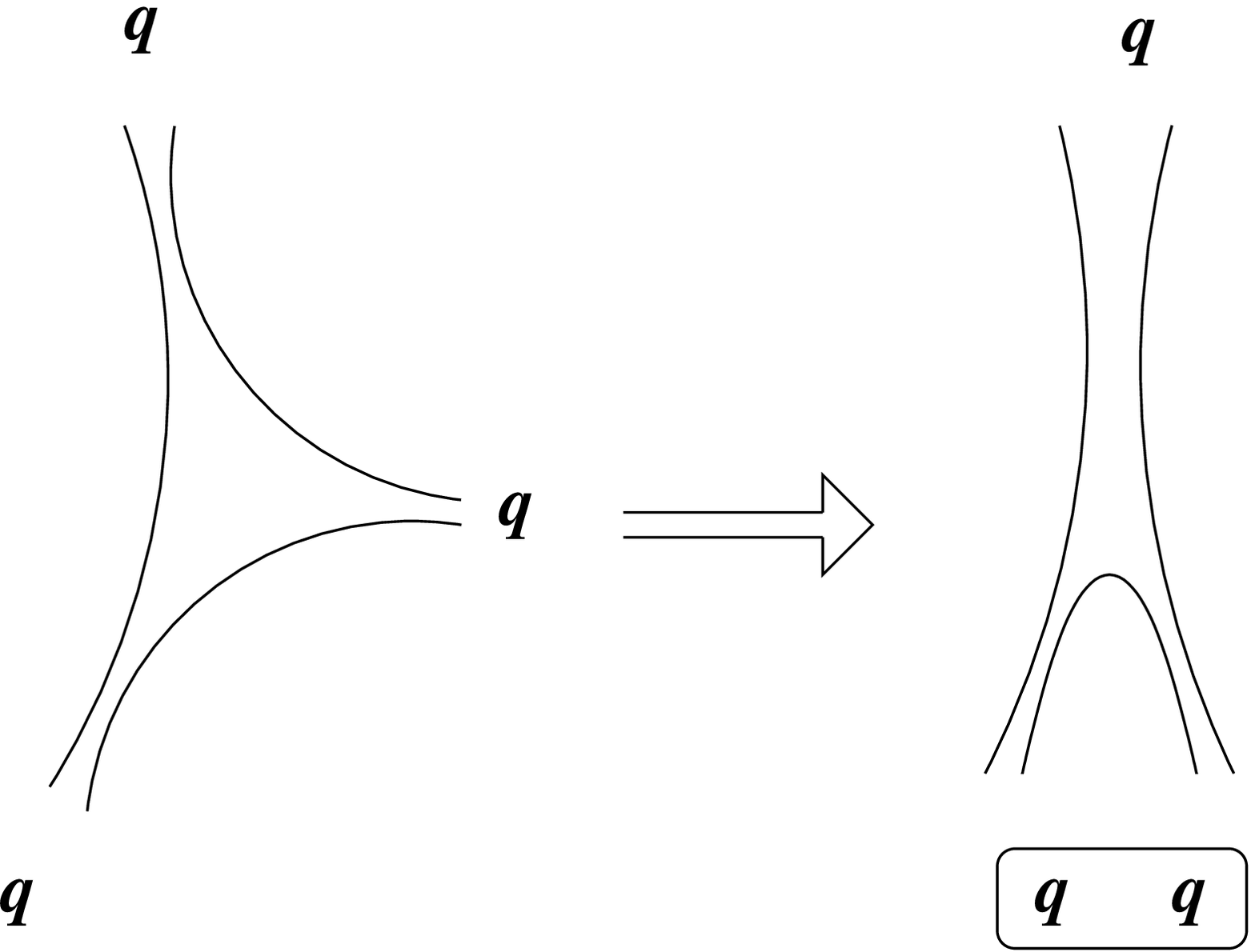,width=5cm}}
\vspace{0.1cm}
{\small Fig.~2.
\hspace{0.2cm} Shrinkage of a brane to a string.
}
\label{fig:brane_string}
\end{figure}

\vspace{0.15cm}
\noindent
Within the context of the quark--di-quark (q-Dq) 
model, the ideas of the vibron model, known from the spectroscopy of 
diatomic molecules \cite{diat} acquires importance as 
a tool for the description
of the excitations of the (q-Dq) vibrator.

\subsection{The quark vibrator.}

In the rigid vibrator model (RVM) the relative (q-Dq) motion
is described by means of four types of boson creation
operators $s^+, p^+_1, p^+_0$, and $p^+_{-1}$
(see Refs.~\cite{diat}, \cite{KiMoSmi}).. 
The operators $s^+$ and $p^+_m$ in turn 
transform as rank-0, and rank-1 spherical tensors,
i.e. the magnetic quantum number $m$ takes
in turn the values $m=1$, $0$, and $-1$.
In order to construct boson-annihilation operators that
also transform as spherical tensors, one introduces
the four operators $\widetilde{s}=s$, and
$\widetilde{p}_m=(-1)^m\, p_{-m}$.
Constructing rank-$k$ tensor product of
any rank-$k_1$ and rank-$k_2$ tensors, say, $A^{k_1}_{m_1}$ 
and $A^{k_2}_{m_2}$, is standard and given by
\begin{equation}
\lbrack A^{k_1}\otimes A^{k_2}\rbrack^k_m =
\sum_{m_1,m_2}(k_1m_1 k_2m_2\vert km)\, A^{k_1}_{m_1}A^{k_2}_{m_2}\, .
\label{clebsh}
\end{equation}
Here, $(k_1m_1k_2m_2\vert km)$ are the standard $O(3)$ Clebsch-Gordan
coefficients.

Now, the lowest states of the two-body system are identified with $N $
boson states and are characterized by the ket-vectors 
$\vert n_s\, n_p\, l\, m\rangle $ (or, a linear combination of them)
within a properly defined Fock space. The constant  
$N=n_s +n_p$ stands for the total number of $s$- and $p$ bosons
and plays the r\'ole  of a parameter of the theory.
In molecular physics, the parameter $N$ is usually associated
with the number of molecular bound states.
The group symmetry of the vibrator is well known to be $U(4)$. 
The fifteen generators of the associated $su(4)$ algebra 
are determined as the following set of bilinears 
\begin{eqnarray}
A_{00}=s^+ \widetilde{s}\, , &\quad& A_{0m}=s^+\widetilde{p}_m\, , 
\nonumber\\
A_{m0}=p^+_m\widetilde{s}\, , &\quad & A_{mm'}=p^+ _m
\widetilde{p}_{m'}\, .
\label{RVM_u4}
\end{eqnarray} 
The $u(4)$ algebra is then recovered by the following 
commutation relations
\begin{equation}
\lbrack A_{\alpha\beta},A_{\gamma\delta}\rbrack_-=
\delta_{\beta \gamma}A_{\alpha\delta}-
\delta_{\alpha\delta}A_{\gamma\beta}\, .
\end{equation}
The operators associated with physical observables can then be expressed
as combinations of the $u(4)$ generators.
To be specific, the three-dimensional angular momentum takes the form
\begin{equation}
L_m=\sqrt{2}\, \lbrack p^+ \otimes \widetilde{p}\rbrack^1_m \, .
\label{a_m}
\end{equation}
Further operators are  $(D_m$)-- and $(D'_m$) defined as 
\begin{eqnarray}
D_m &=&\lbrack p^+\otimes \widetilde{s}+s^+\otimes 
\widetilde{p}\rbrack^1_m\, ,\\
\label{x_dipol_rvm}
D_m '&=&i\lbrack p^+\otimes \widetilde{s}-s^+\otimes 
\widetilde{p}\rbrack^1_m\, ,
\label{p_dipol_rvm}
\end{eqnarray}
respectively.
Here, $\vec{D}\, $ plays the r\'ole of
the electric dipole operator.

{}Finally, a quadrupole operator $Q_m$ can be constructed as
\begin{equation}
Q_m=\lbrack p^+\otimes \widetilde{p}\rbrack^2_m\, ,
\quad \mbox{with}\quad m=-2,..., +2\, .
\label{quadr_rvm}
\end{equation}

\subsection{Degeneracy of the rigid vibrator.}

The $u(4)$ algebra has the two algebras $su(3)$, and $so(4)$, as 
respective sub-algebras. The $so(4)$ sub-algebra of interest here,
is constituted by the three components of the angular momentum operator 
$L_m$, on the one side,
and the three components of the operator $D_m'\, $, on the other side. 
The chain of reducing $U(4)$ down to $O(2)$  
\begin{eqnarray}
&&U(4)\supset O(4)\supset O(3)\supset O(2)\, ,\nonumber\\
&&N\qquad\quad K\qquad\quad  l\qquad\quad m\, , 
\label{chains}
\end{eqnarray}
corresponds to an exactly soluble rigid vibrator model (RVM) limit.
In the second row of Eq.~(\ref{chains}) we indicate the quantum
numbers associated with the respective group of the chain.
Here, $N$ stands for the principle quantum number of the four dimensional
harmonic oscillator associated with $U(4)$,
$K$ refers to the four dimensional angular momentum in $O(4)$,
while $l$, and $m$ are in turn ordinary three-- and two angular momenta.
 
\vspace{0.05cm}
\noindent
In order to demonstrate how this model applies to baryon spectroscopy,
let us consider the case of (q-Dq) states associated with $N=5$. 
It is of common
knowledge that the totally symmetric irreps of the $u(4)$ algebra 
with the Young scheme $\lbrack N\rbrack$ contain the 
$SO(4)$ irreps $\left({K\over 2}, {K\over 2}\right)$ with
\begin{equation}
K=N, N-2, ..., 1 \quad \mbox{or}\quad 0\, .
\label{Sprung_K}
\end{equation}
Each one of the $K$- irreps contains $SO(3)$ multiplets with
three dimensional angular momentum
\begin{equation}
l=K, K-1, K-2, ..., 1, 0\, .
\label{O3_states}
\end{equation}
In applying the branching rules in Eqs.~(\ref{Sprung_K}),
(\ref{O3_states})
to the case $N=5$, one encounters the three series of levels
\begin{eqnarray}
K&=&1: \quad l=0,1;\nonumber\\
K&=&3: \quad l=0,1,2,3;\nonumber\\
K&=&5: \quad l=0,1,2,3,4,5\, .
\label{Ns_Ks}
\end{eqnarray}
The parity carried by these levels is $\eta (-1)^{l}$ where
$\eta $ is the parity of the di-quark. In coupling now the
angular momentum in Eq.~(\ref{Ns_Ks}) to the spin-${1\over 2}$ of the three
quarks in the nucleon, the following sequence of states is obtained:
\begin{eqnarray}
K&=&1: \quad \eta J^\pi={1\over 2}^+,{1\over 2}^-, {3\over 2}^-\, ;
\nonumber\\
K&=&3: 
\quad \eta J^\pi={1\over 2}^+,{1\over 2}^-, {3\over 2}^-,
{3\over 2}^+, {5\over 2}^+, {5\over 2}^-, {7\over 2}^- \, ;
\nonumber\\
K&=&5: \quad \eta J^\pi={1\over 2}^+,{1\over 2}^-, {3\over 2}^-,
{3\over 2}^+, {5\over 2}^+, {5\over 2}^-, {7\over 2}^- ,
{7\over 2}^+, {9\over 2}^-, {11\over 2}^-\, .
\label{set_q}
\end{eqnarray}
Therefore, states of half-integer spin emerging from the
underlying vibrator modes transform according to  
$\left( {K\over 2},{K\over 2}\right) \otimes \left[
\left({1\over 2},0 \right) \oplus 
\left( 0,{1\over 2} \right)\right] $
representations of $SO(4)$, a result due to \cite{KiMoSmi}.
There, we accounted pragmatically for isospin structure through
attaching to the $K$--mega states an isospin spinor
$\chi^I$ with $I$ taking the values $I={1\over 2}$ 
and $I={3\over 2}$ for the nucleon, and the $\Delta $ states,
respectively.
Notice, that as long as compositness of the di--quark has been
completely ignored here (the di-quark is
rather treated as a fundamental scalar/pseudoscalar),
the question about anti--symmetrization of the baryon wave
function does not stay.

\vspace{0.15cm}
\noindent
As illustrated by Fig.~1, the above quantum numbers completely cover
both the nucleon and the $\Delta $ excitations.
The states in  Eq.~(\ref{set_q}) are degenerate and the dynamical symmetry
is $O(4)$. 

\vspace{0.15cm}
\noindent
Above considerations apply to the rest frame.
In order to describe mega states in flight one needs to
subject the O(4) degenerate resonance
states to a Lorentz boost. 
One can  achieve this task either
boosting them  spin by spin, or boosting the $K$-mega state as
a whole. In favoring the latter option one gains
the possibility to map (up to form factors)  $K$-mega states onto
totally symmetric rank-$K$ Lorentz tensors with Dirac spinor components,
$\psi_{\mu_1...\mu_K}$.\footnote{See Ref.~\cite{PLB_02} 
for details on a totally neutral $\psi_{\mu}$, associated with the
massive gravitino, and Ref.~\cite{FF_03} for a discussion on 
why $\psi_{\mu_1...\mu_K}$ (for any K) qualify also for 
the description of charged baryon mega states.}

\subsection{  Observed and ''missing'' $K$ mega states.}

The comparison of the states in Eq.~(\ref{set_q}) with the reported ones
in  Fig.~1 shows that the predicted quantum numbers
of the resonance series (with masses below $\sim $ 2500 MeV) 
would be in agreement with the quantum numbers of the reported 
non-strange baryon excitations  
only if the rigid di-quark changes from scalar ($\eta =1$) for the $K=1$, 
to pseudoscalar ($\eta =-1$) for the $K=3,5$ series.
That is why we assign unnatural parities to the resonances
from the second and third $K$ mega states.

A pseudoscalar di-quark can be modeled in terms of
an excited  composite  di-quark carrying an internal 
angular momentum  $L=1^-$ and maximal spin $S=1$. In one of
the possibilities the total spin of such
a system can be $\vert L-S\vert = 0^-$.

\vspace{0.15cm}
\noindent
To explain the ground state, one has to
consider separately even $N$ values, such as, say, $N'=4$.
In that case another branch of excitations, with $K=4$, $2$, and
$0$ will emerge. The $K=0$ value characterizes the 
ground state,  $K=2$ corresponds to
$\left( 1,1\right)\otimes 
\lbrack\left( {1\over 2},0\right)\oplus 
\left(0,{1\over 2}\right) \rbrack $, while  $K=4$ corresponds to
$\left( 2,2\right) \otimes\lbrack\left( {1\over 2},0\right)\oplus 
\left(0,{1\over 2}\right) \rbrack $. 
These are the multiplets that we 
shall associate with the  ``missing'' resonances  
predicted by the rigid vibrator model.
In this manner, reported and ``missing'' resonances  fall apart 
and populate  distinct $U(4)$--, and therefore distinct $SO(4)$ 
representations.
In making observed and ``missing'' resonances distinguishable,
reasons for their absence or, presence in the spectra are
easier to be searched for.
In accordance with Ref.~\cite{MK_2000}
we here will treat the $N=4$ states to be all of natural parities
and identify them with the nucleon $(K=0)$, the natural parity $K=2$,
and  the natural parity $K=4$-mega states.
We shall refer to the latter as ''missing'' $K$-mega states.

\vspace{0.15cm}
\noindent
The RVM  Hamiltonian is constructed in the standard way
as a properly chosen function of the Casimir 
operators of the algebras of the subgroups entering the chain. 
{}For example, in case one approaches $O(3)$ via $O(4)$,
the algebraic Hamiltonian of a dynamical $SO(4)$ symmetry can be cast into 
the form \cite{KiMoSmi}:
\begin{equation}
H_{RVM}=H_0 - f_1\, \left(4 {\cal C}_2\left( so(4)\right) +1\right)^{-1}
+f_2\, {\cal C}_2(so(4) )\, .
\label{H_QRVM}
\end{equation}
The Casimir operator ${\cal C}_2\left(so(4)\right)$ is defined accordingly as
\begin{equation}
{\cal C}_2\left( so(4)\right)={1\over 4}\left( \vec{L}\, ^2 + 
\vec{D}\, ' \, ^2
\right)\, 
\label{so(4)_Casimir}
\end{equation}
and has an eigenvalue of ${K\over 2}\left( {K\over 2}+1 \right)$.

\vspace{0.15cm}
\noindent
The parameters of the Hamiltonian that fits masses of the observed 
$K$-mega states are:
\begin{equation}
H_0= M_{N/\Delta } +f_1\, ,\quad  f_1=600\,\, \mbox{MeV}\, , \quad 
f_2^N=70\, \, \mbox{MeV}\, , \quad
f_2^\Delta =40 \, \, \mbox{MeV}\, .
\label{f_s}
\end{equation}
Thus, the $SO(4)$ dynamical symmetry limit of the
RVM picture of baryon structure explains    
observation of (quasi)degenerate resonance series
in both the $N$- and $\Delta $ baryon spectra.
In Table I we list the masses of the $K$--mega states concluded from
Eqs.~(\ref{H_QRVM}), and (\ref{f_s}).

\begin{table}[htbp]
\caption{
Mass distribution of observed (obs), and
missing (miss) vibrator  $K$-mega states (in MeV) concluded
from Eqs.~(9,11). 
The sign of $\eta $ in Eq.~(15) labels natural-
($\eta =+1$), or, unnatural ( $\eta =-1$) parity states.
The measured mass averages of the resonances from a given
$K$-mega state have been denoted by ``exp''.}
\vspace*{0.21truein}
\begin{tabular}{lccccccc}
\hline
~\\
K & sign $\eta $ & N$^{\mbox{obs} }$  & N$^{\mbox{exp}}$ &
 $ \Delta^{\mbox{obs}}$ & $\Delta^{\mbox{exp}}$ &  
   N$ ^{\mbox{miss}}$ & $\Delta^{\mbox{miss}}$  \\
\hline
~\\
0 & + &939 &939 & 1232& 1232 & &  \\
1&+  &1441 & 1498 & 1712 &1690 &         & \\
2&+  &     &      &      &     &  1612   & 1846 \\
3&-  &1764 &1689  & 1944 &1922 &         &     \\
4&+  &     &      &      &     &  1935   & 2048 \\
5&-  &2135 &2102  & 2165 &2276 &         &      \\
\hline
\end{tabular}
\end{table}
In order to translate the Hamiltonian in Eq.~(\ref{f_s}) into coordinate
space,
\beq
{\mathcal H}=-\frac{\hbar^2}{2\mu}\nabla^2 +V(r)\, ,
\eeq
one can consider a central potential $V(r)$ that is 
the following combination of Coulomb-- and Morse--like potentials 
\begin{eqnarray}
 V(r)&=&  -\frac{\alpha^2}{r} + V_0(1-e^{-\beta \, r})^2\, ,
\nonumber\\
\alpha^2= \frac{2f_1}{\mu }\, , &\quad&
f_2=-\frac{\beta^2}{\mu}\, ,\quad
V_0=\frac{ (N+2)^2\, \beta}{8\mu }\, .
\label{Coul_Morse}
\end{eqnarray}
The $\sim 1/r$ potential reproduces the $1/2(K+1)^2$ contribution to
the mass, $M_K$, of the excited $K$ mega states coming from
the $\left(4 {\cal C}_2\left( so(4)\right) +1\right)^{-1}$ term in 
Eq.~(\ref{H_QRVM}), while the Morse potential describes the
$K/2(K/2+1)$ contribution to $M_K$ coming from the
${\cal C}_2(so(4) )\,$ term in Eq.~(\ref{H_QRVM})
(see Ref.~\cite{diat}) for more details on the latter point). 
Here $\mu $ stands for the reduced mass of the quark and the di-quark.
Equation (\ref{Coul_Morse}) provides the ground for the calculation
of the wave RVM wave functions and thereby for a more profound
analysis of the decay properties of the $K$-mega states, a subject that 
is currently under investigation.

\subsection{Electromagnetic gross properties of $K$ mega states.}
The four dimensional Racah algebra that 
allows to calculate transition probabilities for electromagnetic 
de-excitations of the vibrator levels was presented in
Ref.~\cite{KiMoSmi}. The interested reader is invited 
to consult this article for details. Here I restrict myself to 
reporting on the following two results:
\begin{enumerate}
\item All resonances from a $K$-mega state have same widths.
\item As compared to the natural parity $K=1$ states,
the electromagnetic de-excitations of the unnatural parity
$K=3$ and $K=5$ RVM states appear strongly suppressed.  
\end{enumerate}
To see how above predictions match with experiment, 
I compiled in Table 2 below data on the measured total widths of resonances 
belonging to  $K=3$, and $K=5$.

\begin{table}[htbp]
\caption{Measured total widths of $K$-mega states }
\begin{tabular}{lcc}
\hline
~\\
K&Resonance & width [in GeV]  \\
\hline
~\\
3 &$N\left( {1\over 2}^-;1650\right)$ & 0.15  \\
3 &$N\left( {1\over 2}^+;1710\right)$ & 0.10   \\
3 &$N\left( {3\over 2}^+;1720\right)$ & 0.15    \\
3 &$N\left( {3\over 2}^-;1700\right)$ & 0.15     \\
3 &$N\left( {5\over 2}^-;1675\right)$ & 0.15      \\
3 &$N\left( {5\over 2}^+;1680\right)$ & 0.13       \\
~\\
\hline
5 &$N\left( {3\over 2}^+;1900\right)$ & 0.50\\
5 &$N\left( {5\over 2}^+;2000\right)$ &0.49\\
~\\\hline
\end{tabular}
\end{table}
Above Table is supportive of the first prediction.
The predicted suppression of the  electromagnetic de--excitation modes 
of unnatural parity states to the nucleon (of natural parity) is due to
the vanishing overlap between the scalar di-quark  in the latter case, 
and the  pseudo-scalar one, in the former. In Table III we list
some available data on electromagnetic helicity amplitudes of resonances.
The Table shows that data are not too much on variety with the predicted 
suppression.
\begin{table}[htbp]
\caption{ Measured  electromagnetic helicity amplitudes of resonances }
\begin{tabular}{lcccc}
\hline
~\\
K& parity of the spin-0 di-quark  
&Resonance  &$A^p_{{1\over 2}}$ & $A_{{3\over 2}}^p$ 
[in $10^{-3}$GeV$^{-{1\over 2}}$]  \\
\hline
~\\
3 &- & $N\left( {1\over 2}^+;1710\right)$  &  9 $\pm$22 &  \\
3 &- & $N\left( {3\over 2}^+;1720\right)$  &  18$\pm$30 &-19$\pm$20   \\
3 &- & $N\left( {3\over 2}^-;1700\right)$  & -18$\pm$30 & -2$\pm$24    \\
3 &- & $N\left( {5\over 2}^-;1675\right)$  & 19 $\pm$8  & 15$\pm$9      \\
3 &- & $N\left( {5\over 2}^+;1680\right)$  & -15$\pm$6  &133$\pm$12      \\
~\\
\hline
1 &+ &$N\left( {3\over 2}^-;1520\right) $ & -24$\pm$9 & 166$\pm$ 5\\
~\\\hline
\end{tabular}
\end{table}
In order to give an interpretation of significantly non-vanishing 
electromagnetic widths within the advocated model,  
one may entertain the idea that an enhanced helicity amplitude
may signal small admixtures from natural parity states of same spins
belonging to mega states of even $K$ that belong to
the ``missing'' modes.
{}For example, the significant value of $A_{{3\over 2}}^p$ for 
$N\left( {5\over 2}^+;1680\right)$
from $K=3$ may appear as an effect of mixing with the 
edge N$\left( {5\over 2}^+;1612\right)$ state
from the natural parity ``missing'' mega states with $K=2$.
This gives one the idea to
use helicity amplitudes to extract ``missing'' states.

\section{Conclusions}
The degeneracy phenomenon in the spectra of the light quark 
baryons was successfully explained in terms of $K$-mega states,
i.e. $\left(\frac{K}{2},\frac{K}{2}\right)\otimes 
\left[\left(\frac{1}{2},0\right)\oplus\left(0,\frac{1}{2}\right)\right]$,
taking their origin from the  quark version of the rigid vibrator 
with a dynamical O(4) symmetry.
We predicted series of degenerate ``missing'' resonances with $K$ even.
To conclude, the $N$, or, $\Delta$ excitation modes seem to fit 
better into  ply-spectra composed by nearly equidistant $K$-mega states
as displayed in Fig.\ 3, but into any other patterns.
To account for the relatively small crumbling of the mega states one 
needs to weaken the dominance of the quark--(rigid di--quark) 
configuration and consider mixing  with three quark states.
It is an intriguing task to build up baryon spectra from
the rigid vibrator and upon subjecting it to small symmetry violations,
to account for the realistic mass distribution of baryon resonances.
The advocated model captures better but any other
quark model the essentials of QCD string solutions.

\vspace*{1truecm}
\begin{figure}[htb]
\vskip 5.0cm
\includegraphics{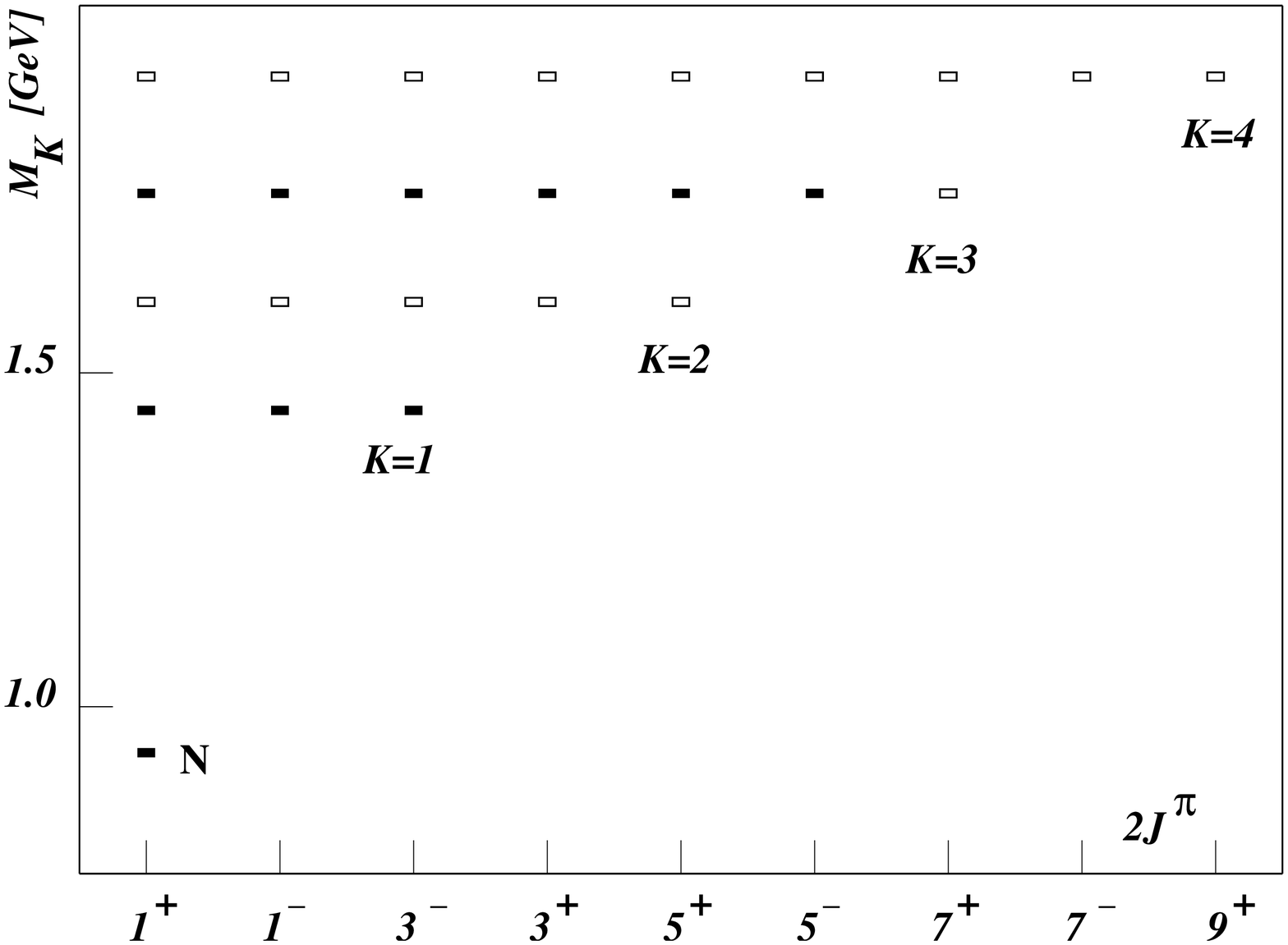}
\includegraphics{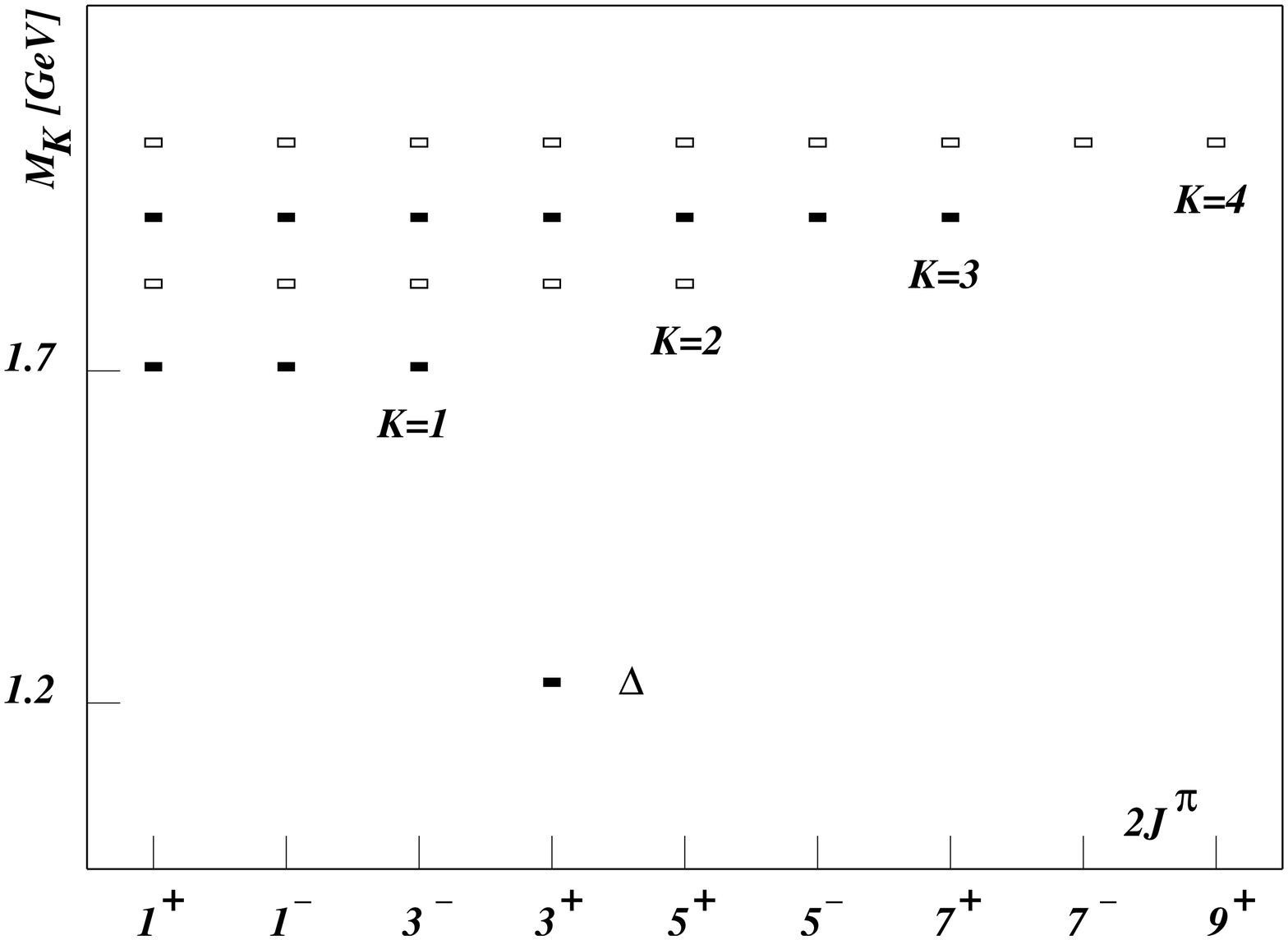}
\vspace{1.01cm}
{\small Fig.~3.\hspace{0.2cm}$N$ and $\Delta$ ply-spectra 
(below 2000 MeV) following from the quark--(rigid di-quark) vibrator. 
Full bricks correspond to observed, empty bricks to ``missing'' resonances.
Comparison to Fig.~1 shows that above model (i) reproduces precisely 
the K-mega states quantum numbers, and specifically those of the edge 
states, (ii) the degeneracy symmetry, i.e the dominance of the rigid di--quark
is best pronounces around $\sim 1700$ MeV for the nucleon and
around $\sim 1900$ MeV  for the $\Delta$, while below and above
the symmetry seems to suffer some violation, provided, the observed
splittings are not artifacts of data analysis.
Data requires natural parities for $K$=1, and unnatural
for $K=3$, and $K=5$ (the latter not shown on the Figure).
The parity for the ``missing'' $K=2$, and $4$ states has
been suggests as natural on the basis of a comparison with
shell model quantum numbers (see Ref.~\cite{KiMoSmi} for details).
Compared to the $N$ spectrum, the  $\Delta $ spectrum, in following 
same patterns, appears somewhat tighter due to the different value we used 
for the parameter $f_2$ in Eq.~(18).  }
\end{figure}

\section*{Acknowledgments}
Collaboration with Marcos Moshinsky and Yuri Smirnov
on the quark--(di-quark) dynamics behind the $K$-mega states
is highly appreciated.

Work supported by Consejo Nacional de Ciencia y
Tecnolog\'ia (CONACyT, Mexico) under grant number 
C01-39820.

\end{document}